\documentclass[reprint,amssymb,aps,superscriptaddress]{revtex4-1}
\usepackage{graphicx}
\usepackage{dcolumn}
\usepackage{bm}
\usepackage{url}
\usepackage{lipsum}
\usepackage{amsmath}
\usepackage{hyperref}
\usepackage[mathlines]{lineno}

\newcommand{\al}{a_{\rm L}}
\newcommand{\bl}{b_{\rm L}}
\newcommand{\kt}{k_{\rm T}}
\newcommand{\ncoll}{n_{\rm coll}}
\newcommand{\pt}{p_{\rm T}}
\newcommand{\raa}{R_{\rm AA}}
\newcommand{\rpa}{R_{\rm pA}}
\newcommand{\rppb}{R_{\rm pPb}}
\newcommand{\siglq}{\sigma_{\rm LQ}}
\newcommand{\sighq}{\sigma_{\rm HQ}}

\begin{document}
\title{Resolving the $R_{\rm pA}$ and $v_2$ puzzle of $D^0$ mesons in $p-$Pb collisions at the LHC}

\author{Chao Zhang}
\affiliation{School of Science, Wuhan University of Technology, Wuhan, 430070, China}
\affiliation{Department of Physics, East Carolina University, Greenville, NC 27858, USA}
\affiliation{Institute of Particle Physics and Key Laboratory of Quark\&Lepton Physics (MOE),\\
Central China Normal University, Wuhan 430079, China}
\author{Liang Zheng}
\affiliation{School of Mathematics and Physics, China University of Geosciences (Wuhan), Wuhan 430074, China}
\author{Shusu Shi}
\affiliation{Institute of Particle Physics and Key Laboratory of Quark\&Lepton Physics (MOE),\\
Central China Normal University, Wuhan 430079, China}
\author{Zi-Wei Lin}
\email{linz@ecu.edu}
\affiliation{Department of Physics, East Carolina University, Greenville, NC 27858, USA}

\begin{abstract}
It has been a challenge to understand the experimental data on both the nuclear modification factor and elliptic flow of $D^0$ mesons in $p-$Pb collisions at LHC energies. In this work, we study these collisions with an improved multi-phase transport model. By including the Cronin effect (i.e., transverse momentum broadening) and independent fragmentation for charm quarks, we provide the first simultaneous description of the $D^0$ meson $R_{\rm pA}$  and $v_2$ data at $p_{\rm T} \leq 8$ GeV$/c$. The model also reasonably describes the $D^0$ meson $p_{\rm T}$ spectra and the low-$p_{\rm T}$ charged hadron spectra, $R_{\rm pA}$ and $v_2$. Our results show that both parton interactions and the Cronin effect are important for the $D^0$ meson $R_{\rm pA}$, while parton interactions are mostly  responsible for the $D^0$ meson $v_2$. It is thus essential to include the Cronin effect for the simultaneous description of the $D^0$ meson $R_{\rm pA}$ and $v_2$. This work implies that the Cronin effect could also be important for heavy hadrons in large systems. 
\end{abstract}

\maketitle

\textit{Introduction.---}
Heavy flavor hadrons are one of the most  important tools to study the perturbative Quantum Chromo-Dynamics (pQCD) in high energy hadronic collisions~\cite{vanHees:2005wb,Brambilla:2010cs,Andronic:2015wma}. Over  the last two decades, experiments from the Relativistic Heavy Ion Collider (RHIC)  and the Large hadron  Collider (LHC)~\cite{Gyulassy:2004zy,Adams:2005dq,Adcox:2004mh,Muller:2012zq} have collected many data supporting the formation of a hot and dense matter called the quark-gluon-plasma (QGP), and a main goal of high energy heavy ion physics is to study the QGP properties. Heavy quarks provide us a great probe because the heavy  quark mass is much larger than the temperature of the dense matter; therefore, heavy flavor particles may only partially thermalize~\cite{Moore:2004tg} and thus better remember the  interaction history with the medium. 

Two observables are often measured for heavy flavors in heavy ion collisions: the nuclear modification factor $\raa$~\cite{STAR:2014wif,Sirunyan:2017xss,Acharya:2018hre,Adam:2018inb,ALICE:2021rxa,ALICE:2021bib,ALICE:2021kfc} and the elliptic flow $v_2$~\cite{ALICE:2013olq,Adamczyk:2017xur,ALICE:2017pbx,Sirunyan:2017plt,Acharya:2017qps,ALICE:2020pvw,ALICE:2020iug}. 
Several theoretical models, including the Fokker-Planck approach~\cite{Das:2010tj,He:2012df,Cao:2015hia,Lang:2016jpe} and the relativistic Boltzmann transport approach~\cite{Fochler:2010wn,Djordjevic:2013xoa,Xu:2015bbz,Song:2015ykw,Cao:2016gvr}, have been developed to study the nuclear suppression and collective  flows of heavy flavor hadrons at RHIC and LHC. 
It has been realized that $\raa$ and $v_2$ are sensitive to the temperature- and energy-dependence of transport properties of the QGP such as the heavy quark  diffusion and drag coefficients~\cite{Das:2015ana,Rapp:2018qla}. 
They are also sensitive to the hadronization mechanisms including 
quark coalescence and fragmentation~\cite{Lin:2003jy,Greco:2003vf,Oh:2009zj,He:2011qa,Cao:2013ita,Song:2015sfa}. 
Several approaches have shown reasonable agreements with the existing data in  large collision systems, suggesting that charm quarks may flow well with the QGP  medium due to their frequent interactions with the hot and dense matter~\cite{Scardina:2017ipo,Cao:2018ews,He:2019vgs}. 

Similar measurements of heavy flavor mesons have also been made for small systems like $d+$Au collisions at RHIC and $p-$Pb collisions at LHC in recent years~\cite{STAR:2004ocv,Adare:2013lkk,Abelev:2014hha,LHCb:2017yua,Sirunyan:2018toe,CMS:2018duw,Acharya:2019mno,ALICE:2020wla}. 
Little to no nuclear suppression $\rpa$ but a sizable elliptic flow $v_2$ 
have been observed for $D^0$ mesons in $p-$Pb collisions at the LHC energies, 
which has posed a big challenge to theoretical models. 
One expects that a sizable $v_2$ comes from significant interactions of charm quarks with the QGP medium, in either hydrodynamics-based models or parton/hadron transport models. 
On the other hand, a significant interaction of charm quarks with the QGP is expected to inevitably suppress high-$\pt$ charm  hadrons~\cite{Xu:2015iha,Du:2018wsj,Cao:2020wlm}, in contrast to the observed  $D^0$ $\rpa$ being almost flat around the value of unity.  

Some theoretical studies can reproduce either the heavy meson $\rpa$ data~\cite{Fujii:2013yja,Kang:2014hha,Beraudo:2015wsd,Liu:2019lac,Tripathy:2020edo,Santos:2022eee,Ke:2022gkq} or $v_2$ data~\cite{Zhang:2019dth,Zhang:2020ayy}. 
For example, the POWLANG model~\cite{Beraudo:2015wsd} can describe the heavy flavor $\rpa$ but predicts a small charm $v_2$. 
PQCD calculations that consider cold nuclear medium effects are generally able  to describe the charm $\rpa$ data~\cite{Fujii:2013yja,Kang:2014hha,ALICE:2016yta,Liu:2019lac}, 
and so is another pQCD model with a parametrized $\kt$ broadening~\cite{Tripathy:2020edo}. 
Regarding the heavy flavor elliptic flow, the color glass condensate framework 
can describe the charm and bottom $v_2$ in $p-$Pb collisions at LHC~\cite{Zhang:2019dth,Zhang:2020ayy}, which indicates the relevance of  initial-state correlation for heavy quarks in small systems. 
So far, however, there has not been a simultaneous description of both $\rpa$ and $v_2$ of heavy hadrons. In this study, we investigate the $D^0$ meson $\rpa$ and $v_2$ in $p-$Pb collisions at LHC energies with an improved version of a multi-phase transport (AMPT) model.  

\textit{Methods.---} 
The AMPT model~\cite{Lin:2004en,Lin:2021mdn} is a transport model designed  to describe the evolution of the dense matter produced in heavy ion collisions. The string melting version~\cite{Lin:2001zk} is expected to be applicable when the QGP is formed, as it contains a fluctuated initial condition, partonic scatterings, quark coalescence, and hadronic interactions.  Recently, we have developed a new quark coalescence~\cite{He:2017tla}, incorporated modern parton distribution functions of the free proton 
and impact parameter-dependent nuclear shadowing~\cite{Zhang:2019utb},  improved heavy flavor productions~\cite{Zheng:2019alz}, 
and applied local nuclear scaling to two input parameters~\cite{Zhang:2021vvp}. 
The AMPT model that we use in this study contains these improvements. 

In the string melting version of AMPT model, the excited strings are converted to partons through the string melting mechanism~\cite{Lin:2001zk}. In particular, the strings are first converted to hadrons through the Lund string fragmentation~\cite{Andersson:1983ia,Andersson:1983jt}, then each hadron is decomposed to partons according to the flavor and spin structures of its valence quarks. Because initial charm quarks are produced from hard pQCD processes during the primary nuclear-nuclear collision, we improve their treatment in this work. Instead of ``melting'' the initial charm hadrons into charm quarks via string melting, we extract charm quarks produced from the HIJING model~\cite{Wang:1991hta} before they enter the Lund string fragmentation. These initial charm quarks then enter the parton cascade; and a charm quark is allowed to interact after its formation time given by $t_{\rm F}=E/m_{\rm T}^2$~\cite{Lin:2004en}, where $E$ and $m_{\rm T}$ are the quark energy and transverse mass, respectively. 

Since the scattering cross section for charm quarks is in general different from that for light ($u,d,s$) quarks, we separate the cross section among light quarks ($\siglq$) from that between a heavy quark and other quarks ($\sighq$). 
The default values, $\siglq=0.5$ mb and $\sighq=1.5$ mb, are used unless specified otherwise, and they are determined from the fit to the charged hadron  
$v_2$ data in $p-$Pb collisions at 5.02 TeV and $D^0$ meson $v_2$ data in $p-$Pb collisions at 8.16 TeV, respectively. We have also added the independent fragmentation~\cite{Sjostrand:1993yb}  as another hadronization process for  heavy quarks, in addition to the usual quark coalescence process~\cite{He:2017tla}. If a heavy quark and its coalescing partner(s) have a large relative distance or a  large invariant mass, they are considered to be unsuitable for quark coalescence; instead the heavy quark will hadronize to a heavy hadron via independent fragmentation.

We also include the transverse momentum broadening (i.e., the Cronin effect~\cite{Cronin:1974zm}) for the initial heavy quarks~\cite{Mangano:1992kq,Vogt:2018oje}.  
The Cronin effect is often considered as the broadening of the transverse momentum of a produced parton from a hard process 
due to multiple scatterings of the involved parton(s) in the nucleus ~\cite{Kopeliovich:2002yh,Kharzeev:2003wz,Vitev:2006bi,Accardi:2002ik}. 
Therefore, its strength depends on the number of scatterings a participant (or target) nucleon undergoes while passing the target (or projectile)  nucleus~\cite{Vogt:2021vsc}. 
We implement the broadening by adding a transverse momentum kick $\kt$ to each $c\bar{c}$ pair in the initial state, where $\kt$ is sampled from a two-dimensional Gaussian~\cite{Mangano:1992kq,Vogt:2018oje,Vogt:2021vsc}
with a Gaussian width parameter $w$:
\begin{eqnarray}
&&f (\vec \kt)=\frac{1}{\pi w^2} e^{-\kt^2/w^2},\\
\label{fkt}
&&w=w_0 \sqrt{1+(\ncoll-i) \delta}.
\label{width}
\end{eqnarray}
Note that a $c\bar{c}$ pair can be produced from either the radiation of one participant nucleon or the collision between one participant nucleon from the projectile and another from the target. In Eq.\eqref{width}, $i=1$ for the former case and $i=2$ for the latter case, while $\ncoll$ is the number of primary NN collisions of the participant nucleon for the former case and the sum of the numbers  of primary NN collisions of both participant nucleons for the latter case. 
This way, $w=w_0$ for $p{+}p$ collisions, where 
\begin{equation}
w_0=(0.35{\rm~GeV}/c)~\sqrt {\bl^0(2+\al^0)/\bl/(2+\al)}.
\end{equation}
In the above, $\al^0=0.5$ and $\bl^0=0.9$ GeV$^{-2}$ are the original values in  the HIJING1.0 model~\cite{Wang:1991hta} for the two parameters in the Lund fragmentation function~\cite{Sjostrand:1993yb}, and $\al$ and $\bl$ are the values in the AMPT model~\cite{Zhang:2021vvp}. 
The dependence of $w_0$ on the Lund parameters is based on the observation that the average squared transverse momentum of a hadron relative to the fragmenting parent string is proportional to the string tension, which  scales as $1/\bl/(2+\al)$~\cite{Lin:2004en}. 
We take $\al=0.8$ and determine $\bl$ according to the local  nuclear thickness functions, where the $\bl$ value is $0.7$ GeV$^{-2}$ for $p{+}p$ collisions but smaller for nuclear collisions~\cite{Zhang:2021vvp}. As a result, for $p{+}p$ collisions, $w=0.375$ GeV$/c$, close to the original  value of $0.35$ GeV$/c$ for the parameter parj(21) in the HIJING1.0 model~\cite{Wang:1991hta}.  The $\delta$ in Eq.\eqref{width} controls the strength of the Cronin effect; its default value of $\delta=5.0$ is determined from the comparison with the $D^0$ meson $\rpa$ data.

In the implementation of the Cronin effect, we give each $c\bar{c}$ pair a transverse boost so that the pair transverse momentum increases by a  $\vec \kt$ sampled from the distribution in Eq.\eqref{fkt}. 
Note that such implementation of the Cronin effect tends to 
create an artificial peak at mid-rapidity in the rapidity distribution of heavy quarks~\cite{Vogt:2018oje,Vogt:2019xmm,Vogt:2021vsc}, 
since $y={\rm arcsinh}(p_{\rm z}/m_{\rm T})$ will move towards zero after $\pt$ increases. Therefore, we choose to keep the rapidity of $c\bar{c}$ pair the same by providing the necessary longitudinal boost after the transverse momentum broadening. We also enforce the momentum conservation of the whole parton system of each event by letting the light (anti)quarks share the opposite value of the total $\vec \kt$ given to all $c\bar{c}$ pairs in the event. 

\begin{figure}[!htb]
\includegraphics[scale=0.43]{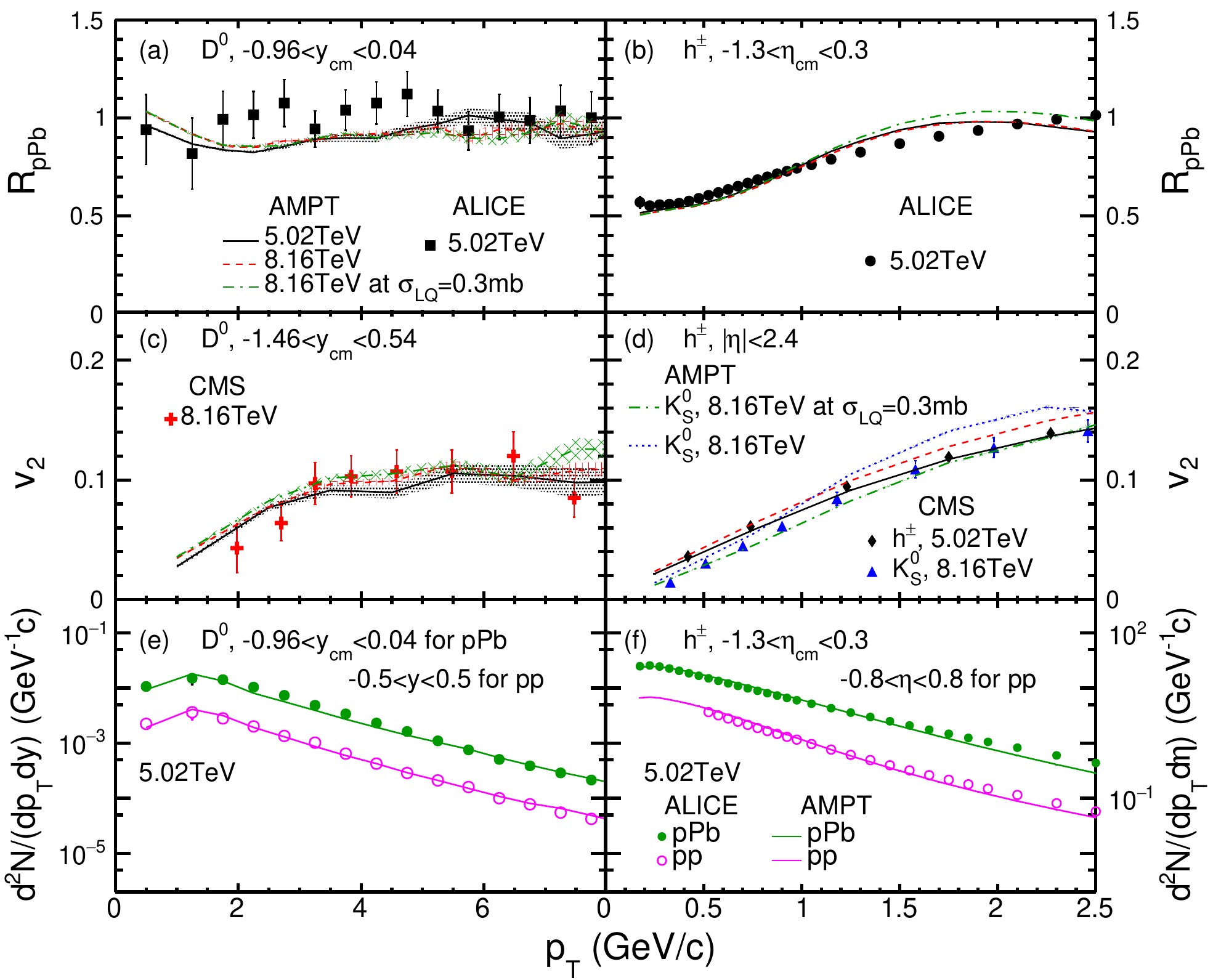}
\caption{$\rppb$ of (a) $D^0$ mesons and (b) charged hadrons
in minimum bias $p-$Pb collisions, $v_2$ of (c) $D^0$ mesons, (d) charged hadrons and $K_{\rm S}^0$ in high multiplicity $p-$Pb collisions, and the $\pt$ spectra  of (e) $D^0$ mesons and (f) charged hadrons in minimum bias $p-$Pb and $p{+}p$ collisions at 5.02 TeV from the improved AMPT model in comparison with the experimental data around mid-rapidity.}
\label{fig1}
\end{figure}

\textit{Results and discussions.---}
Figure~\ref{fig1} shows in the upper panels 
our results of the nuclear modification factor  $\rppb$ 
as functions of the transverse momentum for $D^0$ mesons and charged hadrons in minimum bias $p-$Pb collisions at 5.02 TeV and 8.16 TeV in comparison with the experimental data. The middle panels show the elliptic flow  coefficient $v_2\{2\}$ in high multiplicity $p-$Pb collisions. 
All results in Fig.~\ref{fig1} are obtained with the full AMPT model, with $\siglq=0.5$ mb (except for the dot-dashed curves where $\siglq=0.3$ mb), $\sighq=1.5$  mb, and $\delta=5.0$. We see from panels (a) and (c) that this AMPT model can simultaneously describe the available $D^0$ meson $\rppb$ data at 5.02 TeV~\cite{Acharya:2019mno} and $v_2$ data at 8.16 TeV~\cite{Sirunyan:2018toe} below $\pt \sim 8$ GeV$/c$. In addition, as shown  in panels (b) and (d), the model  well describes the charged hadron $\rppb$~\cite{Acharya:2018qsh} and  $v_2$~\cite{Chatrchyan:2013nka} at 5.02 TeV (solid curves) and reasonably describes the $K_{\rm S}^0$ $v_2$ at 8.16 TeV~\cite{Sirunyan:2018toe} below $\pt \sim 1$ GeV$/c$. Furthermore, panels (e) and (f) show the $D^0$ meson and charged hadron $\pt$ spectra in minimum bias $p-$Pb and $p{+}p$ collisions at 5.02 TeV. We see that the AMPT model can well describe the $D^0$ $\pt$ spectra data~\cite{Acharya:2019mno} in both $p{+}p$ and $p-$Pb systems, while the agreements with the charged hadron $\pt$ spectra data~\cite{Acharya:2018qsh}
are reasonable below $\pt \sim 1.5$ GeV$/c$. 

In our analysis, we follow the exact procedures of the ALICE and CMS experiments~\cite{Chatrchyan:2013nka,Sirunyan:2018toe,Acharya:2018qsh,Acharya:2019mno}. Specifically, the $D^0$ meson and charged hadron nuclear modification factors are analyzed for minimum bias collisions within $-0.96<y_{\rm cm}<0.04$ and $-1.3<\eta_{\rm cm}<0.3$, respectively. The elliptic flow coefficient is analyzed for high multiplicity $p-$Pb  events within $N_{\rm track} \in  [185-220)$ at 5.02 TeV and $N_{\rm track} \in  [185-250)$ at 8.16 TeV, where $N_{\rm track}$ is the number of charged hadrons with $\pt>0.4$ GeV$/c$   within $|\eta|<2.4$. 
To calculate the elliptic flow from two-particle correlations, we apply  $|\Delta\eta|>2$ at 5.02 TeV and $|\Delta\eta|>1$ at 8.16 TeV, where charged hadrons are selected within $|\eta|<2.4$ while $D^0$ and $K_{\rm S}^0$ mesons are within  $-1.46<y_{\rm cm}<0.54$. The elliptic flow $v_2\{2\}$, written as $v_2$ for brevity, is calculated as~\cite{Chatrchyan:2013nka,Sirunyan:2018toe}
\begin{equation}
v_2(\rm tri)=V_{2\Delta}({\rm tri,ref})/\sqrt {V_{2\Delta}({\rm ref,ref})}, 
\label{v2c}
\end{equation}
where ``tri'' represents the trigger particle of interest, and ``ref'' represents a reference charged hadron with $0.3<\pt<3.0$ GeV$/c$.  
Note that in this study the result of a particle species represents the average of the particle and its corresponding anti-particle; also, all the rapidity and $\eta$ cuts refer to their values in $p-$Pb (not Pb$-p$) collisions.

Since the available data on $D^0$ mesons are the $\rppb$ at 5.02 TeV and $v_2$ at 8.16 TeV, we also show in Fig.~\ref{fig1}(a) and (c) the predictions of $\rppb$ at 8.16 TeV (dashed curve) and $v_2$ at 5.02 TeV (solid curve). 
We see that the $\rppb$ results are almost the same at the two energies but $v_2$ shows an increase with the colliding energy. This is also the case for the  charged hadron $\rppb$ and $v_2$, as shown by the dashed curves for 8.16 TeV in Fig.~\ref{fig1}(b) and (d). We note that the model overestimates the  $v_2$ of  $K_{\rm S}^0$ mesons at 8.16 TeV when $\siglq=0.5$ mb, which well reproduces the charged hadron $v_2$ at 5.02 TeV, is used. On the other hand, the parton scattering cross section $\sigma$ could be different at different energies. For example, the shear viscosity-to-entropy ratio satisfies $\eta/s \propto 1/(n^{2/3} \sigma)$ for a parton gas in equilibrium under isotropic scatterings~\cite{Huovinen:2008te,MacKay:2022uxo}, where $n$ is the parton number density. As a result, for a constant $\eta/s$, $\sigma$ would be smaller at higher densities. For anisotropic scatterings, which is the case for the AMPT model, the relationship between $\eta/s$ and $\sigma$ is more complicated but qualitatively similar~\cite{MacKay:2022uxo}. 
Therefore, we have also explored the effect of a different light quark cross section. As shown by the dot-dashed curves in Fig.~\ref{fig1}(a)-(d), changing $\siglq$ from 0.5 mb to 0.3 mb at 8.16 TeV enables the AMPT model to well reproduce the $K_{\rm S}^0$ $v_2$ data, but this change has almost no effect on the $D^0$ meson $\rppb$ and $v_2$. As expected, the smaller $\siglq$ leads to a small enhancement of the charged hadron $\rppb$, as shown in Fig.~\ref{fig1}(b). 

\begin{figure}[!htb]
\includegraphics[scale=0.43]{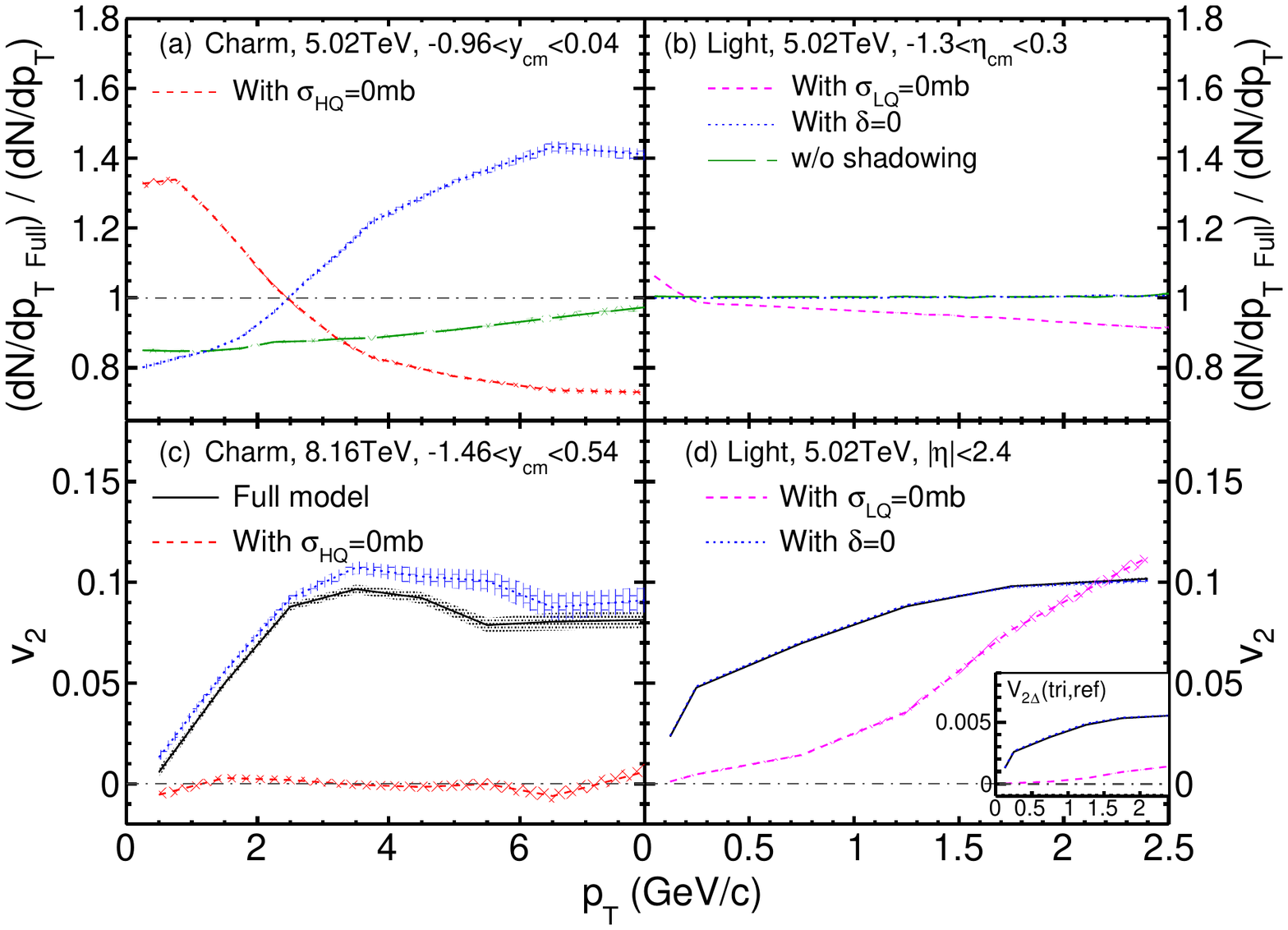}
\caption{Ratio of the $\pt$ spectrum from the full AMPT model over that from
the AMPT model with a different test configuration for (a) charm quarks and (b) light  quarks in $p-$Pb collisions at 5.02 TeV, (c) $v_2$ of charm quarks at 8.16 TeV, and (d) $v_2$ of light quarks at 5.02 TeV from the AMPT model for $p-$Pb collisions. The Cronin effect is turned off with $\delta=0$. The inset in panel   (d) shows the light quark ${\rm V}_{2\Delta}({\rm tri,ref})$.}
\label{fig2}
\end{figure}

We now separately turn off various effects to identify the key ingredients that allow the improved AMPT model to simultaneously describe the $D^0$ meson $\rppb$ and $v_2$. Figure~\ref{fig2}(a) shows the ratio of the charm quark $\pt$ spectrum from
the full AMPT model over that from different test configurations of the AMPT model for minimum bias $p-$Pb collisions at 5.02 TeV, while Fig.~\ref{fig2}(b)  shows the ratios for light quarks. 
The dashed curves in panels (a) and (c) represent the results of charm quarks  without charm quark scatterings (but with scatterings among light quarks),  while the dashed curves in panel (b) and (d) represent the light quark results without scatterings among light quarks. 
We see that parton scatterings suppress the parton yield at relatively high $\pt$ (and enhance the yield at low $\pt$) due to the parton energy loss or jet quenching~\cite{Ozvenchuk:2017ojj,Xu:2015iha}; this effect is especially significant  for charm quarks, partially due to the larger scattering cross section for charm quarks. 
From the dotted curves that correspond to turning off the charm Cronin effect, 
we find that the Cronin effect significantly enhances the charm quark yield at relatively high $\pt$ and essentially cancels out the effect from jet quenching. 
In addition, we see that the EPS09s nuclear shadowing~\cite{Zhang:2019utb}
has almost no effect on the light quark $\pt$ spectrum but 
a modest suppression effect on the charm quark $\pt$ spectrum in the transverse momentum range shown in Fig.~\ref{fig2}.

We show in Fig.~\ref{fig2}(c) and (d) the results on the charm quark $v_2$ at 8.16 TeV and the light quark $v_2$ at 5.02 TeV, respectively, for the high multiplicity  $p-$Pb collisions. From the dashed curves, we see that the charm quark $v_2$ is mostly generated from the scatterings of charm quarks with the medium, while the initial state correlation before rescatterings (or non-flow) contributes significantly to the light  quark $v_2$ but little to the charm quark $v_2$. We also see that the Cronin effect for charm quarks modestly suppresses the charm quark $v_2$; it has little effect on the light quark $v_2$, as expected. Note that in Fig.~\ref{fig2}(d) the light quark $v_2(\pt)$ without scatterings among light quarks is even higher than that with parton scatterings at $\pt > 2.2$ GeV$/c$. The inset in Fig.~\ref{fig2}(d) shows the corresponding numerator, ${\rm V}_{2\Delta}({\rm tri,ref})$, for the light quark $v_2$, where the result without scatterings is significantly lower than that with parton scatterings, as expected. Therefore, the relatively high $v_2(\pt)$ without scatterings is due to the fact that the denominator  $\sqrt {{\rm V}_{2\Delta}({\rm ref,ref})}$ in Eq.\eqref{v2c}, which corresponds to the reference elliptic flow, is much smaller without scatterings. 

\begin{figure}[!htb]
\includegraphics[scale=0.43]{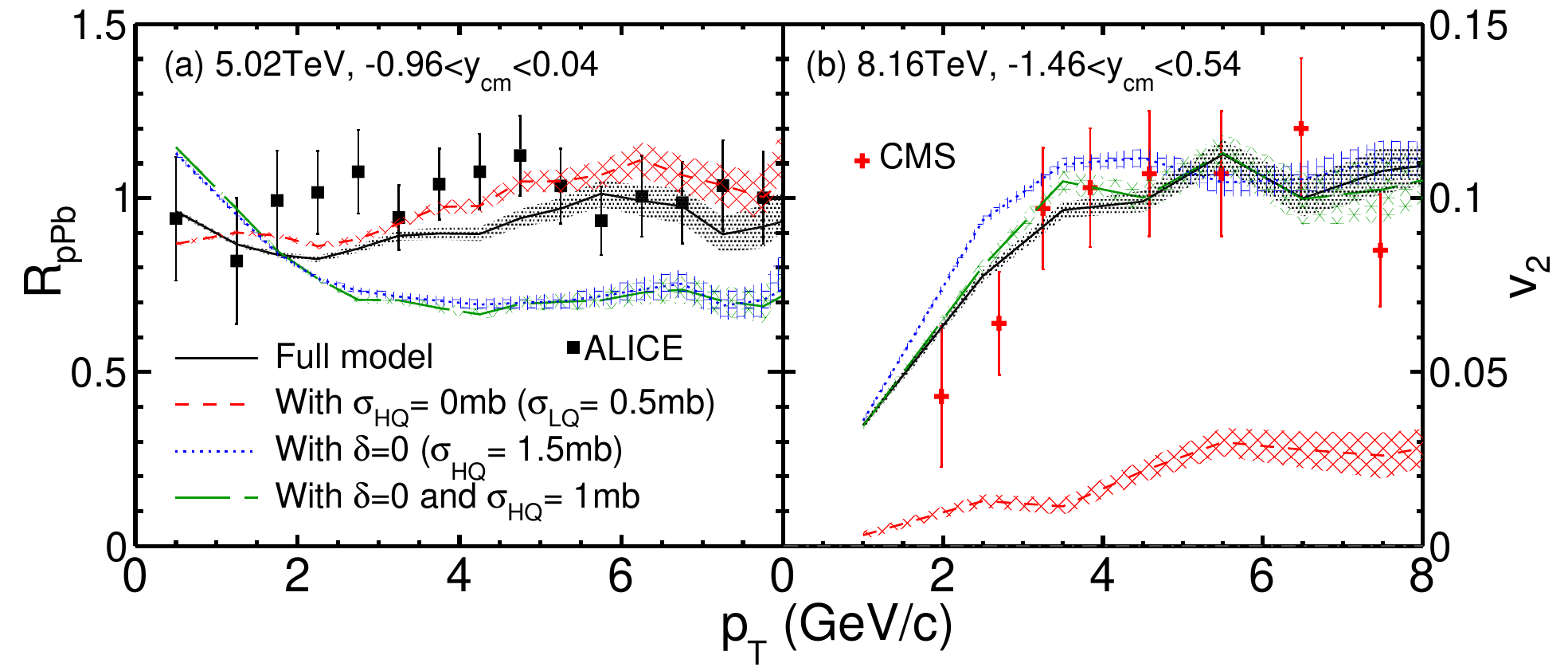}
\caption{(a) $\rppb$ at 5.02 TeV and (b) $v_2$ at 8.16 TeV for $D^0$ mesons in $p-$Pb collisions from the full AMPT model (solid), the model without charm quark scatterings (dashed), the model without the Cronin effect for charm quarks (dotted), and the model without the Cronin effect at a smaller charm  quark scattering cross section (long-dashed) in comparison with the experimental data  (symbols).}
\label{fig3}
\end{figure}

We now examine the effects of transverse momentum broadening and parton scatterings on the $D^0$ meson $\rppb$ and $v_2$. 
When the Cronin effect is turned off (with $\delta=0$), we see in Fig.~\ref{fig3}(a) that the $D^0$ $\rppb$ is significantly suppressed at high  $\pt$ but enhanced at low $\pt$. Therefore, the Cronin effect is very important for the  $D^0$ meson $\rppb$. 
In addition, parton scatterings (at $\sighq=1.5$ mb) are seen to suppress the  $D^0$ meson $\rppb$ at high $\pt$, qualitatively the same as its effect on  charm quarks as shown in Fig.~\ref{fig2}(a). 
Quantitatively, the effect of parton scatterings on the $D^0$ meson $\rppb$ is 
smaller than that on charm quarks; this is because the fraction of charm quarks hadronizing via quark coalescence (instead of fragmentation) 
increases with the amount of scatterings and consequently the system size. When  charm quark scatterings are turned off (dashed curve) in the AMPT model, the charm quark yield at high $\pt$ is enhanced due to the absence of energy loss. On the other hand, more charm quarks hadronize via independent fragmentation (than the case with charm quark scatterings), which reduces the enhancement of $D^0$ mesons at high $\pt$. 

In Fig.~\ref{fig3}(b), the $D^0$ meson $v_2$ is mostly very small 
when charm quark scatterings are turned off (dashed curve); the $D^0$ $v_2$ is thus mostly generated by parton scatterings, similar to the charm quark $v_2$ shown in Fig.~\ref{fig2}(c). 
Note that, even if charm quarks have zero $v_2$, the $D^0$ $v_2$ can be finite since it has a contribution from the light quark $v_2$ through quark coalescence~\cite{Lin:2003jy}. 
In the AMPT model without the Cronin effect, the $D^0$ meson $v_2$ (dotted curve) is slightly higher. Therefore, the Cronin effect modestly suppresses the $D^0$ $v_2$. 
We have also decreased the charm quark scattering cross section to 1.0 mb, from the default value of 1.5 mb in the full model, to better fit the $D^0$ meson $v_2$ (long-dashed curve). 
The corresponding $D^0$ meson $\rpa$ result is shown in Fig.~\ref{fig3}(a) as the long-dashed curve, which is seen to still severely underestimates the data at  high $\pt$. The Cronin effect is thus crucial for the simultaneous description of the $D^0$ meson $\rppb$ and $v_2$ data according our model calculations.

Many previous theoretical methods and phenomenological models 
have found the Cronin effect to be important. For example, pQCD results~\cite{Mangano:1992kq} have indicated that the Cronin effect is needed to describe the experiment data of open heavy flavors at fixed-target energies.  
In the pQCD-based HVQMNR code~\cite{Vogt:2018oje,Vogt:2021vsc}, 
transverse momentum broadening is also needed to describe quarkonium $\pt$ distributions and heavy flavor azimuthal distributions from fixed-target to LHC energies. In the HVQMNR code, a transverse momentum kick in the form of Eq.\eqref{fkt} is applied to each produced heavy quark in $p{+}p$ collisions, where  the Gaussian width is energy-dependent~\cite{Vogt:2018oje}:$\langle \kt^2\rangle_p=[1+\ln(\sqrt{s_{NN}}/20/{\rm GeV})/n]\, \rm GeV^2$ 
with $n{=}12$ for $J/\psi$ productions.
For minimum-bias $p{+}A$ collisions, the Gaussian width increases to 
$\langle\kt^2\rangle_A=\langle\kt^2\rangle_p+\delta\kt^2$~\cite{Vogt:2021vsc}, where  $\delta\kt^2=(1.5\rho_0R_A\sigma_{pp}^{in}-1)\Delta^2(\mu)$. 
Here, $R_A=1.2A^{1/3}\,{\rm fm}$ represents the nuclear radius, $\rho_0=0.16/{\rm fm}^3$ is the average nuclear density, $\sigma_{pp}^{in}$ is the inelastic $p{+}p$ cross section, and $\Delta(\mu)=0.318\,{\rm GeV}$ for charm productions at $\mu=2m_c=2.54\,{\rm GeV}$~\cite{Vogt:2021vsc}.
Note that we apply the broadening to each $c\bar c$ pair, while the HVQMNR code applied it to each charm (anti)quark in the final state~\cite{Vogt:2021pbi}; therefore,  we have calculated the $\kt$ broadening to each charm quark in the comparisons  below. For $p{+}p$ collisions at 5.02 TeV, the HVQMNR code~\cite{Vogt:2021vsc,Vogt:2022glr} gives $\langle \kt^2 \rangle =  1.46\,\rm{GeV}^2$, higher than our value of $0.04\,{\rm GeV}^2$. For minimum bias $p$-Pb collisions at 5.02 TeV, the HVQMNR code gives $\langle \kt^2  \rangle=2.50\,{\rm GeV}^2$, lower than our value of $3.27\,{\rm GeV}^2$. 

The AMPT model currently only includes the collisional energy loss via two-body elastic parton scatterings, while the parton radiative energy loss is not included.  In the relativistic limit, the heavy quark collisional energy loss has been shown to depend on the path length $L$ linearly while the radiative energy loss scales as $L^2$~\cite{Mustafa:2004dr}. 
Therefore, the collisional energy loss of charm quarks is expected to 
be more important than the radiative energy loss for small systems like $p$-Pb, 
although the $\pt$ scale below which the collisional energy loss dominates is 
model-dependent~\cite{Cao:2013ita,Nahrgang:2013saa,Ke:2018jem,Ke:2020clc}. 
In addition, the radiative energy loss of charm quarks through inelastic collisions would suppress the charm $\pt$ spectrum at high $\pt$, qualitatively the same as the collisional energy loss through elastic collisions. 
Therefore, the inclusion of the charm quark radiative energy loss would not change our conclusion that the Cronin effect is needed to compensate for the effect of energy loss and consequently describe the observed $D^0$ meson $\rppb$ and  $v_2$ simultaneously.

Since the Cronin effect is expected to be stronger for a larger collision system, our study also suggests that it would be important to include the Cronin effect in studies of light hadron~\cite{Vitev:2002pf} or heavy hadron $\raa$~\cite{He:2012df} in large systems. Currently, several models are able to reasonably describe the $D$ meson $\raa$  and $v_2$~\cite{Song:2015sfa,Cao:2016gvr,Scardina:2017ipo,He:2019vgs}. 
The inclusion of the Cronin effect may change the model results and affect the extracted values of the charm quark transport coefficients. Therefore, further studies, including those with a predicted (instead of a fit) strength for the Cronin  effect and those on charmonium observables, will lead to a better understanding of the roles of cold nuclear matter and hot medium effects on heavy flavor productions in small to large collision systems.

\textit{Summary.---}
We have studied the $D^0$ meson as well as charged hadron nuclear modification factor $\rppb$ in minimum bias $p-$Pb collisions and elliptic flow $v_2$ in high multiplicity $p-$Pb collisions at LHC energies with a multi-phase transport model. After improving the model with the transverse momentum  broadening (i.e., the Cronin effect) and independent fragmentation for charm quarks, 
we are able to provide the first simultaneous description of both the 
$\rppb$ and $v_2$ data of $D^0$ mesons below the transverse momentum of 8 GeV$/c$. In addition, the transport model reasonably describes the $D^0$ meson  $\pt$ spectra in both $p-$Pb and $p{+}p$ collisions and the low-$\pt$ charged hadron $\pt$  spectra, $\rppb $ and $v_2$. 
We find that  both parton scatterings and the Cronin effect significantly affect the $D^0$ meson $\rppb$. On the other hand, the $D^0$ meson $v_2$ is mostly generated by parton scatterings, while the Cronin effect leads to a modest reduction of the charm $v_2$. In particular, we demonstrate the importance of the Cronin effect for resolving the $D^0$ meson $\rppb$ and $v_2$ puzzle at LHC energies. 
Since the Cronin effect is expected to grow with the system size, this study also implies the importance of including the Cronin effect in studies of heavy hadron $\raa$ and $v_2$ in large systems.

\textit{Acknowledgement ---}
We thank Jacek Otwinowski for the clarification about the ALICE trigger. 
This work is supported by the National Key Research and Development Program of China under contract Nos. 2022YFA1604900 and 2020YFE0202002 (C.Z. and S.S.), 
the National Natural Science Foundation of China under Grant Nos. 12175084, 11890710 (11890711) (C.Z. and S.S.) and 11905188 (L.Z.), the Chinese Scholarship Council (C.Z.), and the National Science Foundation under Grant No. 2012947 (Z.-W.L.). ZWL thanks the Institute for Nuclear Theory at the University of Washington for its kind hospitality and discussions with Ramona Vogt and Peter Petreczky during the revision of this work.

\bibliography{reference}

\end{document}